\documentclass{PoS-hep}
\usepackage{amsmath}
\usepackage{bm}
\usepackage{epsfig}
\usepackage{graphics}
\usepackage{upgreek}
\usepackage{amsfonts}
\usepackage{amsbsy}
\usepackage{amscd}
\usepackage{bbm}
\usepackage{eufrak}
\usepackage{cite}

\renewenvironment{subequations}{%
\refstepcounter{equation}%
\setcounter{parentequation}{\value{equation}}%
  \setcounter{equation}{0}
  \ignorespaces
}{%
  \setcounter{equation}{\value{parentequation}}%
  \ignorespacesafterend
}
\newcommand{\eeq}{\end{equation}}
\newcommand{\beq}{\begin{equation}}
\newcommand{\ba}{\begin{array}}
\newcommand{\ea}{\end{array}}
\newcommand{\bea}{\begin{eqnarray}}
\newcommand{\eea}{\end{eqnarray}}
\newcommand{\baq}{\begin{eqnarray}}
\newcommand{\eaq}{\end{eqnarray}}

\newcommand{\ecs}{\end{cases}}
\newcommand{\bcs}{\begin{cases}}

\newcommand{\beqs}{\begin{subequations}}
\newcommand{\eeqs}{\end{subequations}}
\newcommand{\eec}{\end{center}}
\newcommand{\bec}{\begin{center}}
\newcommand{\eem}{\end{matrix}}
\newcommand{\bem}{\begin{matrix}}
\newcommand{\Eref}[1]{Eq.~(\ref{#1})}
\newcommand{\Sref}[1]{Sec.~\ref{#1}}
\newcommand{\Fref}[1]{Fig.~\ref{#1}}
\newcommand{\Tref}[1]{Table~\ref{#1}}
\newcommand{\cref}[1]{Ref.~\cite{#1}}

\newcommand{\etal}{{\it et al.\/}}

\newcommand\eqs[2]{Eqs.~(\ref{#1}) and (\ref{#2})}

\newcommand\eqss[3]{Eqs.~(\ref{#1}), (\ref{#2}) and (\ref{#3})}

\newcommand{\sFref}[2]{Fig.~\ref{#1}-{\small\sf ({#2})}}

\newcommand{\ftn}{\footnotesize}

\newcommand{\GeV}{{\mbox{\rm GeV}}}

\def\lf{\left(}
\def\rg{\right)}

\newcommand{\Vjhi}{\ensuremath{V_{\rm I}}}
\newcommand{\Vhi}{\ensuremath{V_{\rm I}}}
\newcommand{\Hhi}{\ensuremath{H_{\rm I}}}

\newcommand{\Khi}{\ensuremath{K}}

\newcommand{\Vf}{\ensuremath{V_{\rm F}}}
\newcommand{\Vhio}{\ensuremath{V_{\rm I0}}}

\newcommand{\mP}{\ensuremath{m_{\rm P}}}

\def\openone{\leavevmode\hbox{\small1\kern-3.8pt\normalsize1}}
\newcommand{\dV}{\ensuremath{\Delta V_{\rm I}}}
\newcommand{\Dex}{\ensuremath{\Delta_{\star}}}

\newcommand\vevi[1]{\langle {#1} \rangle_{\rm I}}

\newcommand{\ks}{\ensuremath{k_\star}}
\newcommand{\Ns}{\ensuremath{{N_\star}}}
\newcommand{\ns}{\ensuremath{n_{\rm s}}}

\newcommand{\as}{\ensuremath{a_{\rm s}}}
\newcommand{\As}{\ensuremath{A_{\rm s}}}

\newcommand{\rcc}{\ensuremath{\mathcal{R}}}

\newcommand{\Ve}{\ensuremath{V}}

\newcommand{\Whi}{\ensuremath{W_{\rm B}}}

\def\bbet{{\bar\beta}}
\def\al{{\alpha}}

\def\n{\bar{n}}
\def\m{\bar{m}}

\def\th{{\theta}}

\newcommand{\Trh}{\ensuremath{T_{\rm rh}}}
\newcommand{\sg}{\ensuremath{\phi}}

\newcommand{\ld}{\ensuremath{\lambda}}

\newcommand{\sgx}{\ensuremath{\phi_\star}}
\newcommand{\sgf}{\ensuremath{\phi_{\rm f}}}

\newcommand{\what}{\ensuremath{\widehat}}

\newcommand{\se}{\ensuremath{\widehat{\phi}}}
\newcommand{\sex}{\ensuremath{\widehat{\phi}_\star}}
\newcommand{\sef}{\ensuremath{\widehat{\phi}_{\rm f}}}

\newcommand{\eph}{\ensuremath{ \epsilon}}
\newcommand{\ith}{\ensuremath{ \eta}}

\def\Ka{K\"{a}hler potential}

\def\Km{K\"{a}hler metric}
\def\Kaa{K\"{a}hler~}

\newcommand{\plk}{{\it Planck}}

\newcommand{\phc}{\ensuremath{\Phi}}
\newcommand{\phcb}{\ensuremath{\Phi^*}}

\newcommand{\bdhh}{{\ensuremath{\normalsize I{\kern-2.9pt H}}}}

\def\actc{{\sf\small P-ACT-LB-BK18}}

\newcommand{\nm}{\ensuremath{q_{\rm M}}}
\newcommand{\fr}{\ensuremath{j_{{\rm M}}}}
\newcommand{\frs}{\ensuremath{j_{{\rm M}\star}}}
\newcommand{\far}{\ensuremath{j_{{\rm E}}}}
\newcommand{\fars}{\ensuremath{j_{{\rm E}\star}}}
\newcommand{\fbr}{\ensuremath{j_{{\rm T}}}}

\newcommand{\fm}{\ensuremath{f_{\rm M}}}
\newcommand{\fb}{\ensuremath{f_{\rm T}}}
\newcommand{\fbs}{\ensuremath{f_{\rm T\star}}}
\newcommand{\fes}{\ensuremath{f_{\rm E\star}}}
\newcommand{\fe}{\ensuremath{f_{\rm E}}}
\newcommand{\fms}{\ensuremath{f_{\rm M\star}}}

\newcommand{\kst}{\ensuremath{K_{\rm st}}}

\newcommand{\kle}{\ensuremath{K_{\rm E}}}
\newcommand{\kbe}{\ensuremath{K_{\rm E}}}
\newcommand{\kae}{\ensuremath{K_{\rm E}}}

\newcommand{\klt}{\ensuremath{K_{\rm T}}}

\newcommand{\kat}{\ensuremath{K_{\rm T}}}

\newcommand{\tklmst}{\ensuremath{\widetilde K_{\rm Ms}}}
\newcommand{\tklm}{\ensuremath{\widetilde K_{\rm M}}}
\newcommand{\klm}{\ensuremath{K_{\rm M}}}

\newcommand{\nst}{\ensuremath{N_{\rm st}}}
\newcommand{\nb}{\ensuremath{N_{\rm st}}}
\newcommand{\na}{\ensuremath{N}}

\def\tpmi{{T$_p$MI}}
\def\actc{{\sf\small P-ACT-LB-BK18}}

\def\t{\rm T}
\def\e{\rm E}
\def\m{\rm M}
\def\tmi{{TMI}}
\def\emi{{EMI}}

\newcommand{\etmi}{{E/TMI}}
\def\epmi{{E$_p$MI}}
\newcommand{\etpmi}{{E$_p$/T$_p$MI}}

\title{\boldmath  From $N$- to $(p,N)$-Inflationary Attractors in view of ACT}

\ShortTitle{From $N$- to $(p,N)$-Inflationary Attractors in view
of ACT}

\author{\speaker{C. Pallis}\\
School of Technology,  \\ Aristotle University of Thessaloniki,\\
Thessaloniki, GR-541 24 GREECE\\
       E-mail: \email{kpallis@auth.gr}}

\abstract{We review two types of fractional \Ka s $K$ which
reduce, along the inflationary path, to the form
$N/(1-\sg^{\nm})^{p}$ with $\nm=1$ or $2$ and $0.1\leq p\leq10$.
Their coexistence, within a non-linear sigma model, with chaotic
inflationary potentials of the form $\phi^n$ (where $n=2$ or $4$)
determines, independently from $\nm$ and $n$, a class of
$(p,N)$-inflationary attractors which leads to observables
compatible with the ACT DR6. An implementation of these models in
the context of supergravity can be also achieved by introducing
two chiral superfields and a monomial superpotential, linear with
respect to the inflaton-accompanying field, and supplementing the
$K$'s above with a shift symmetry. Although inflation is attained
for subplanckian inflaton values, the tensor-to-scalar ratio
obtained for certain $N$ values can be possibly observable in the
near future.
\\ \\
{\sl\bfseries Published in}~~{PoS  CORFU {\bf 2025}, 212 (2026)}.
}

\FullConference{Corfu Summer Institute 2025 "School and Workshops
on Elementary Particle Physics and Gravity" (CORFU2025)
24 August - 3 September, 2025\\
Corfu, Greece}

\begin{document}

\section{Introduction}

The models of chaotic inflation adopt the power-law potentials of
the form
\beq V_{\rm I}=\ld^2\sg^n\>\>\> \mbox{or}\>\>\>V_{\rm
I}=\ld^2(\sg^2-M^2)^{n/2}\>\>\>\mbox{for}\>\>\>M\ll\mP=1,
\label{ci} \eeq
which are very common in physics and so it is easy the
identification of the inflaton $\sg$ with a field already present
in the theory. If $\sg$ is canonically normalized, i.e., if
$\sg=\se$ -- where $\se$ is the canonically normalized inflaton --
then the theoretically derived values of the scalar spectral index
$\ns$ and/or tensor-to-scalar ratio $r$ for $n=2$ and $4$ are not
consistent with the observational ones. In particular, for fitted
-- see below -- $\As$ and $\Ns$, we obtain \cite{plin}
\beq \ns\simeq0.968~~\mbox{and}~~r\simeq0.12~~\mbox{for
$n=2$}~~\mbox{or}
~~\ns\simeq0.947~~\mbox{and}~~r\simeq0.28~~\mbox{for $n=4$.} \eeq
On the other hand, the \emph{Data Release 6} from the
\emph{Atacama Cosmology Telescope} ({\sf\small ACT}) combined with
\plk, {\sc Bicep2}/{\slshape Keck Array} and DESI results (in
short \actc\ data) \cite{actin} dictates
\beq \ns=0.9743\pm0.0068,~\as  =
0.0062\pm0.0104~~\mbox{and}~~r\leq0.038\hspace*{0.2cm} \mbox{at
95\% c.l.}\label{data}\eeq

Results compatible with \plk\ \cite{plin} (but not ACT) are
obtained in the models called $N$-attractors \cite{alinde,eno7}
which predict, independently from $n$, $\ns=1-2/\Ns\simeq0.968$
for $\Ns\simeq55$ -- for an updated review see \cref{actreview}.
These models employ specific $\sg-\se$ relations, which can be
derived by non-minimal kinetic terms including poles \cite{pole,
pole1, terada} such as
\beq
\frac{N\dot\sg^2}{2\fm^2}\>\>\mbox{where}\>\>\fm=1-\sg^{\nm}\>\>\mbox{and}\>\>\nm=\bcs1&\mbox{for
$\m=\e$,}\\ 2&\mbox{for $\m=\t$,}\ecs\label{fp}\eeq
which define respectively \emph{E and T-model inflation} (\emi\
and \tmi). Here dot denotes derivation \emph{with respect to}
({\ftn\sf w.r.t}) the cosmic time. Such kinetic mixing can be
derived by some \Ka, which is a real function of complex
variables. E.g, we can use \cite{sor,epole,polec,tmhi}
\beq \label{ket} K=\bcs-2N\ln(1-(\Phi+\Phi^*)/\sqrt{2})&\mbox{for
\emi}
\\ -(N/2)\ln(1-2|\Phi|^2)&\mbox{for \tmi}\ecs ~~\mbox{where}~~ \Phi=\sg e^{i\theta}/\sqrt{2}\eeq
is a (complex) scalar field. Along the direction $\vevi{\theta}=0$
-- where the symbol $\vevi{Q}$ stands for the value of a quantity
$Q$ during inflation -- $\fm$ is found by the \Km\ of the
$\Phi-\Phi^*$ space. I.e.,
\beq \label{fbp} N/2\fm^2=\vevi{\partial_\Phi\partial_{\Phi^*}
K}:=\vevi{K_{\Phi\Phi^*}}. \eeq

We here propose new $K$'s -- first introduced in \cref{actpole} --
which reinforce the pole in $\fm$ with a new exponent $p$
rendering \etmi\ consistent with \actc\ data. The resulting models
are called \emph{E$_p$- and T$_p$-Model inflation} ({\small\sf
\etpmi}) or, collectively, $(p, N)$-attractors. These are first
established in a non-SUSY framework in \Sref{set} and then these
are promoted in the context of SUGRA in \Sref{sugra}. Our results
are exposed in \Sref{ana} and \ref{num} following an approximate
analytic and a more accurate numerical approach respectively.
Finally, we summarize our conclusions in \Sref{con}.

\section{Non-SUSY Framework} \label{set}

Working in the context of a non-linear sigma model we assume that
the kinetic mixing in the $\Phi-\phcb$ space is controlled by a
metric $K_{\phc\phcb}$ which originates from a K\"ahler potential
$K$ according to the generic definition
\beq \label{kdef}
K_{\al\bbet}=\partial_{z^\al}\partial_{z^{*\bbet}}
K>0\>\>\>\mbox{with}\>\>\>K^{\bbet\al}K_{\al\bar
\gamma}=\delta^\bbet_{\bar\gamma},\eeq
where $z^\al$ are complex scalar fields. The relevant lagrangian
terms are written as
\beq\label{action1} {\cal  L} = \sqrt{-\mathfrak{g}}
\left(-\frac{1}{2}\rcc +K_{\phc\phcb}
\partial_\mu \phc\partial^\mu \phcb-
V(\phc)\right), \eeq
where $\mathfrak{g}$ is the determinant of the background
Friedmann-Robertson-Walker metric $g^{\mu\nu}$ with signature
$(+,-,-,-)$, $\rcc$ is the Ricci scalar  and star ($^*$) denotes
complex conjugation. Trying to generate kinetic mixing similar to
that in \Eref{fp} we consider two $K$'s, $\klm$ with $\m=\e$ and
$\t$, defined for $\na>0$, in direct correspondence with those in
\Eref{ket}, as follows
\beq\kae={\na}{\left(1-(\phc+\phcb)/2^{1/2}\right)^{-p}}\>\>\>\mbox{and}\>\>\>
\kat={\na}{\left(1-2|\phc|^2\right)^{-p}} \label{ketp}\eeq
with $\phc+\phc^{*}<\sqrt{2}$ or $|\phc|^2<1/2$ respectively and
$0.1\leq p\leq10$ -- for other alternatives see \cref{actpole}.
Here $\phc$ can be parameterized as in \Eref{ket}. Note that
$\kae$ enjoys a shift symmetry \cite{epole} whereas \kat\ is
invariant under a $U(1)$ symmetry. The corresponding \Kaa metrics
are found to be
\begin{align} \label{mab} K_{\phc\phcb}=pN\cdot\begin{cases}
(p+1)/2{(1-(\phc+\phc^*)/\sqrt{2})^{p+2}}&\mbox{for $K=\kae$,}\\
{2(1+2p|\phc|^2)}/{(1-2|\phc|^2)^{p+2}}&\mbox{for $K=\kat$.}
\end{cases}\end{align}
Both $\klm$ in \Eref{ketp} parameterize hyperbolic \Kaa manifolds
but without constant curvatures as in the cases of \etmi\ -- cf.
\cref{polec}.


The potential $V$ for our models in \Eref{action1} assumes the
form
\beq\label{vsg} V(\phc)=\ld^2|\phc|^{n}+m^2|\phc-\phc^*|^2.\eeq
The last unusual term in $V$ provides the angular mode of $\phc$
in \Eref{fp} $\th$ with mass, as we see below -- cf. \cref{ethi}.
$V$ in \Eref{vsg} could give rise to chaotic inflation since along
the direction
\beq \vevi{\th}=0\>\>\mbox{we obtain}\>\>
\Vhi:=\vevi{V}=\ld^2\sg^n/2^{n/2}, \label{vhi}\eeq
The compatibility with data is obtained thanks to the $\sg-\se$
relation which takes the form
\beq \label{VJe}
{d\se}/{d\sg}=J=\vevi{K_{\phc\phcb}}^{1/2}=\bcs\lf
pN\far/2\sg\fe^{p+2}\rg^{1/2}\>\>&\mbox{for $K=\kle$,}\\
\lf 4 pN\fbr^{2}/2\fb^{p+2}\rg^{1/2}\>\>&\mbox{for $K=\klt$,} \ecs
\eeq
where no summation over the repeated indices \e\ or \t\ is
applied, we retain the definition of $\fm$ in \Eref{fp} and
introduce the auxiliary function
\beq \fr=\bcs (1+p)\sg~~&\mbox{for $\m=\e$,}\\
1+p\sg^2 ~~&\mbox{for $\m=\t$.} \ecs \label{fr} \eeq

\begin{figure}[!t]\vspace*{-.18in}
\hspace*{-.12in}
\begin{minipage}{8in}
\includegraphics[width=60mm,angle=-90]{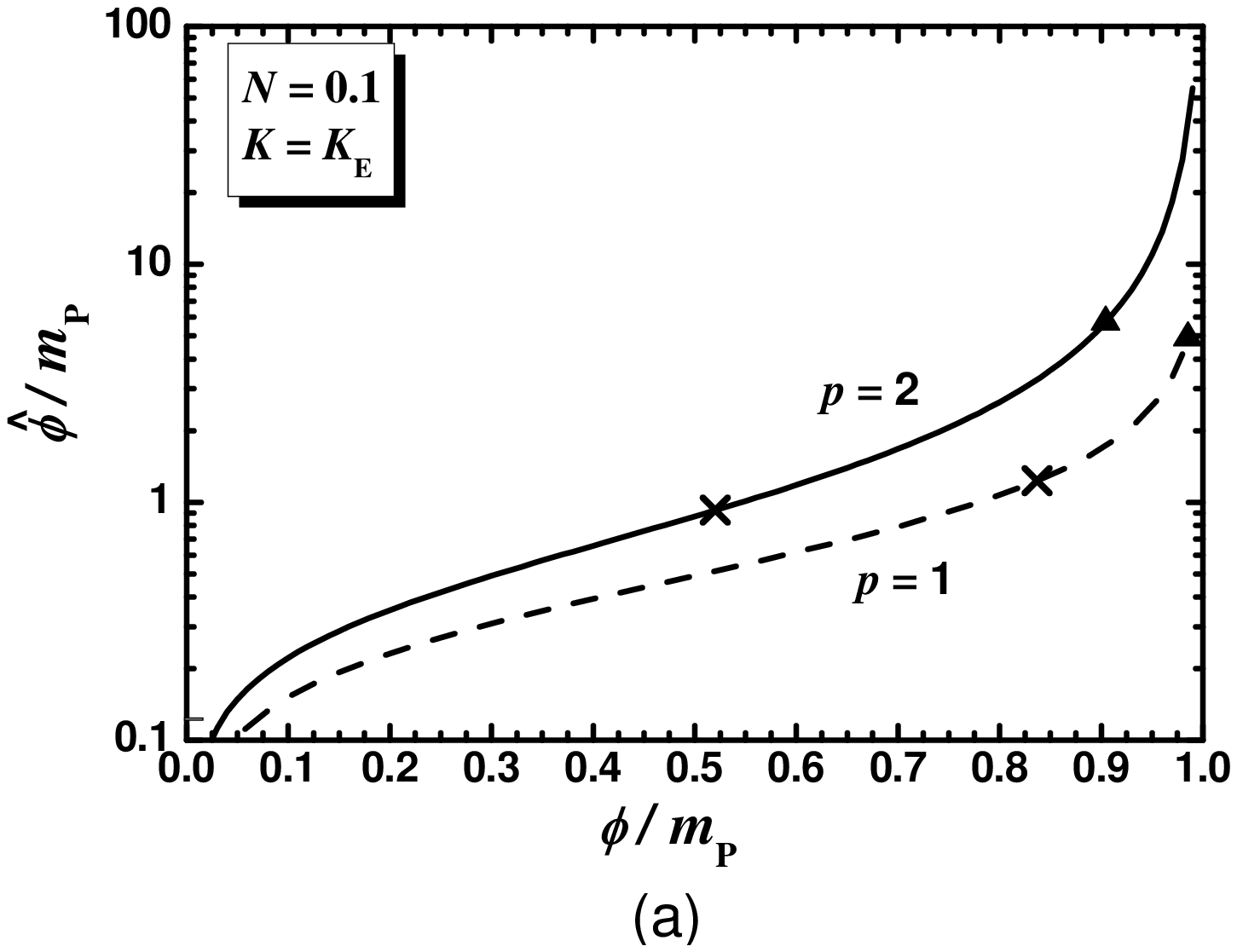}\hspace*{-0.7cm}
\includegraphics[width=60mm,angle=-90]{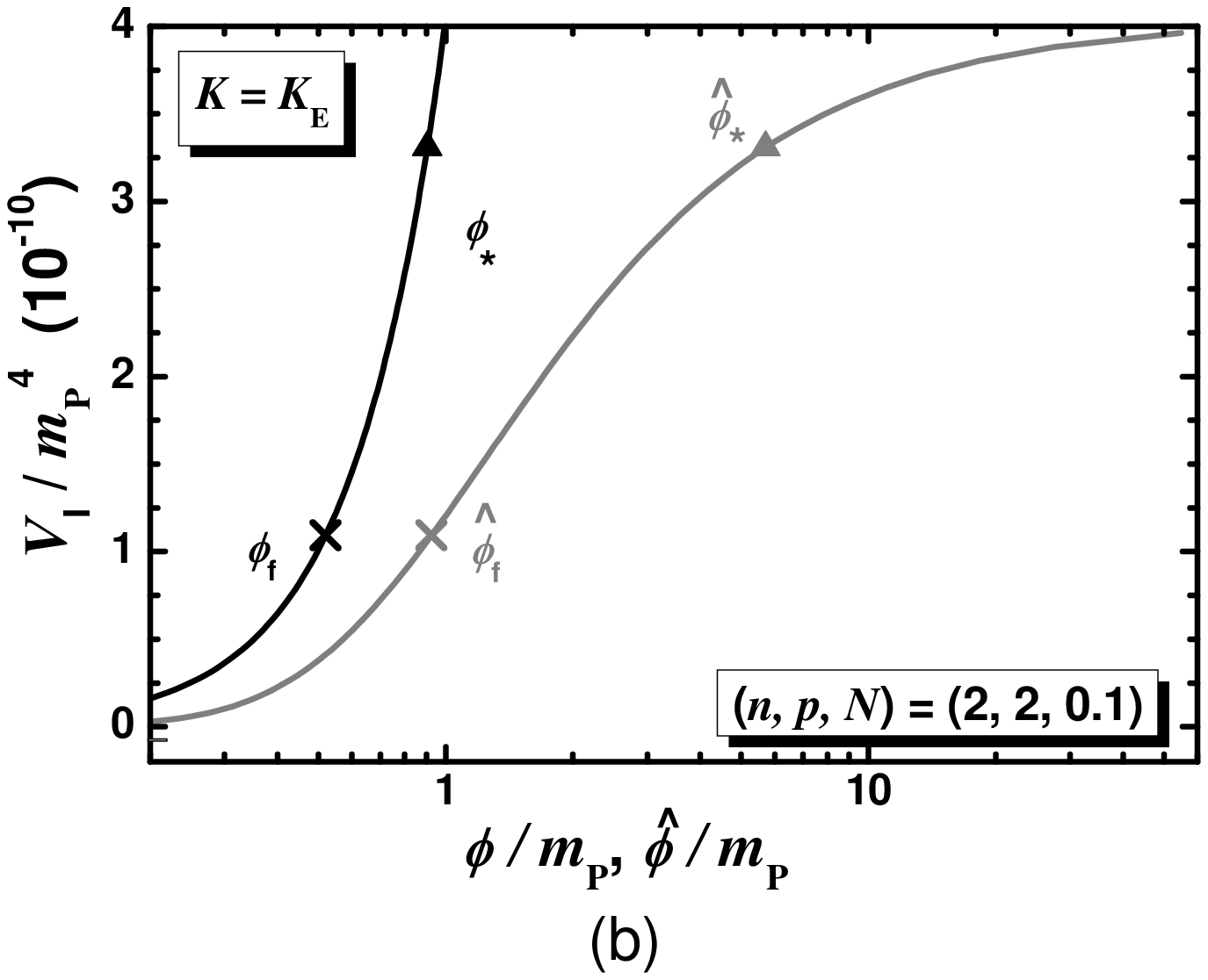}
\end{minipage}
\hfill \caption{\sl\small {\sffamily\small (a)} Canonically
normalized inflaton $\se$ as a function of $\sg$ for \epmi,
$N=0.1$ and $p=2$ (solid line) or $p=1$ (dashed line);
{\sffamily\small (b)} inflationary potential $\Vhi$ for \epmi,
$(n,p,N)=(2,2,0.1)$ as a function of $\sg$ (black line) and $\se$
(gray line). Values corresponding to $\sgx$, $\sgf$, $\sex$ and
$\sef$ are also depicted in both panels.}\label{fig1}
\end{figure}

Integrating \Eref{VJe} we can specify the functions $\se=\se(\sg)$
which have the forms
\beq \se=\sqrt{2 N p}\cdot\bcs \sqrt{(1 + p)/p^2\fe^p}& \mbox{for $K=\kle$,} \\
\sg F_1\lf1/2; 1 + {p}/{2}, -1/2; 3/2; \sg^{2}, -p \sg^2\rg &
\mbox{for $K=\klt$.} \ecs \label{sesg} \eeq
Here $F_1(a;b_1,b_2;c;x,y)$ is the  Appell hypergeometric function
of two variables. To gain a visual understanding of the
expressions above we plot in \sFref{fig1}{a} $\se$ as a function
of $\sg$ for $K=\kbe$, $N=0.1$ and $p=2$ (solid line) or $p=1$
(dashed line). We remark that $\se$ increases beyond unit even for
$\sg<1$ and so \etpmi\ becomes possible since $\Vhi$ for these
$\se$ values develops a plateau, as shown in \sFref{fig1}{b},
where $\Vhi$ is drawn for \epmi, $(n,p,N)=(2,2,0.1)$ as a function
of $\sg$ (black line) and $\se$ (gray line). On the other hand,
$\Vhi$ as a function of $\sg$ has the well-known parabolic-like
slope. This stretching mechanism of $\Vhi$ for $\se>1$ is
well-established within \etmi\ -- see e.g. \cref{alinde,epole,sor}
-- and it remains valid also in our present case. In both panels
\Fref{fig1} we depict also the observationally relevant
inflationary period which is limited between the two $\sg$ values
$\sgf$ and $\sgx$ -- see \Sref{ana} below. Note that although we
set $\mP = 1$ for convenience, we recover it in the axis' labels
of the plots for convenience.


We can, finally, verify that $\what{\th}= J\th\sg$ remains
well-stabilized during \etpmi\ since it acquires heavy mass,
thanks to the last term of \Eref{vsg}. Indeed, we confirm that
during \etpmi
\beq m_\th
=\frac{3\cdot2^{n/2}m^2\fm^{p+2}}{\ld^2pN\fr\sg^n}\Hhi^2\gg\Hhi^2=\frac{\Vhi}{3}\>\>\mbox{for
$K=\klm$,}\label{Hhi}\eeq
provided that $m\gtrsim10^{-2.5}$. Therefore, $\what{\th}$ does
not contribute to the curvature perturbation -- see below in
\Sref{ana}. We also checked that the one-loop radiative
corrections, $\dV$, to $\Vhi$ induced by $m_{\th}$ let intact our
inflationary outputs, if we take for the renormalization-group
mass scale $Q=m_{\th}(\sgx)$ -- cf.~\cref{ethi}.


\section{SUGRA Framework}\label{sugra}

To achieve a SUGRA incarnation of our set-up we consider two
gauge-singlet chiral superfields, i.e., $z^\al=\Phi, S$, with
$\Phi$ ($\al=1$) and $S$ ($\al=2)$ being the inflaton and a
``stabilizer'' field respectively following \cref{rube, su11}. The
relevant part of the  SUGRA Lagrangian density for $z^\al$'s can
be written as
\beqs \beq\label{Saction1}  {\cal  L} = \sqrt{-\mathfrak{g}}
\lf-\frac{1}{2}\rcc +K_{\al\bbet} \partial_\mu z^\al \partial^\mu
z^{*\bbet}-\Vf\rg, \eeq
where summation is taken over the scalar fields $z^\al$ and the
\Ka\ $\Khi$ obeys \Eref{kdef}. Also $\Vf$ is the F--term SUGRA
potential given by
\beq \Vf=e^{\Khi}\left(K^{\al\bbet}(D_\al W) (D^*_\bbet
W^*)-3{\vert W\vert^2}\right),\label{Vsugra} \eeq \eeqs
where $D_\al W=W_{,z^\al} +K_{,z^\al}W$ with $\Whi$ being the
superpotential.

We consider the most general $W$ consistent with the $R$ symmetry
under which $R(S)=R(W)$ and apply some specific hierarchies to
simplify it. Namely,
\beq W=S(\lambda_1\Phi+\lambda_2\Phi^2-M^2)~~\Rightarrow~~W=
\bcs\ld S\phc&\mbox{with}~\ld=\lambda_1~~\mbox{for}~~\lambda_2/\lambda_1\ll0.001\\
\ld
S\phc^{2}&\mbox{with}~\ld=\lambda_2~~\mbox{for}~~\lambda_1/\lambda_2
\ll0.001\ecs \label{Wn}\eeq
and $M\ll\mP$. Therefore, $W$ assumes the following monomial form
\beq  W=\ld S\Phi^{n/2}~~\mbox{with}~~n=2,~4\label{Wn1}\eeq
which may reproduce $\Vhi$ in \Eref{ci} via \Eref{Vsugra} for
suitably selected $K$'s. These same $K$'s have to reproduce the
$\sg-\se$ relation in \Eref{VJe}.

Both objectives can be achieved, as we show below, if we adopt the
the following $K$'s, \tklmst, which include two contributions
without mixing between $\Phi$ and $S$, i.e.,
\beq \tklmst=\tklm+\kst\>\>\mbox{with}\>\>\m=\e,\>\t
\label{ktot}\eeq
where the indices ``s'' and ``st'' are just descriptive, i.e.,
they do not take symbolic values as \m. From the contributions of
$\tklmst$, $\kst$ successfully stabilizes $S$ along the
inflationary path
\beq \label{inftr} \vevi{S}=\vevi{\theta}=0,\eeq
without invoking higher order terms. We adopt the form \cite{su11}
\beq \kst=\nst\ln\lf1+{|S|^2/\nst}\rg\>\>\mbox{with}\>\>0<\nst<6,
\label{kst}\eeq
which parameterizes \cite{su11} the compact manifold $SU(2)/U(1)$
with curvature $2/\nb$. On the other hand, $\tklm$ contains not
only $\klm$ in \Eref{ketp} but also an holomorphic (and an
anti-holomorphic) part which augments it with a shift symmetry
assuring $\vevi{\tklmst}=0$. Namely,
\beqs\beq \tklm=\klm+K_{\rm
Msh}\>\>\mbox{with}\>\>\m=\e,\>\t\label{tks}\eeq
where $K_{\rm Msh}$, with the index ``sh'' being again
descriptive, assumes the form
\beq K_{\rm Msh}=-(\na/2)\left(1-(\sqrt{2}\phc)^{\nm}\right)^{-p}-
(\na/2)\left(1-(\sqrt{2}\phc^{*})^{\nm}\right)^{-p},\label{kmsh}\eeq
which can be specified as follows
\beq K_{\rm Msh}=\bcs -(\na/2)\left(1-\sqrt{2}\phc\right)^{-p}-(\na/2)\left(1-\sqrt{2}\phc^{*}\right)^{-p}\>\>&\mbox{for \m=\e,}\\
-(\na/2)\left(1-2\phc^2\right)^{-p}-(\na/2)\left(1-2\phc^{*2}\right)^{-p}
&\mbox{for \m=\t.}\ecs\label{ksh}\eeq\eeqs
With these ingredients, we can easily confirm that
\beq\vevi{K_{\Phi\Phi^*}}=J^2\>\>\mbox{for}\>\>K=\tklmst\>\>\mbox{with
$J$ given in \Eref{VJe}.}\label{ekss}\eeq

The appropriateness of $W$ and $K$ in \eqs{Wn}{ktot} for the
realization of \etpmi, as described in \Sref{set}, can be verified
if we notice that the only surviving term of $\Vf$ in
\Eref{Vsugra} along the track in \Eref{inftr} is
\beq \label{1Vhio}\vevi{\Vf}=\vevi{e^{K}K^{SS^*}\,
|W_{,S}|^2}\,.\eeq
where the various ingredients can be computed as
\beq\vevi{e^K}=\vevi{K_{SS^*}}=1~~\mbox{and}~~\vevi{|W_{,S}|^2}=\Vjhi\label{2Vhio}\eeq
Therefore, we arrive at $\vevi{\Vf}=\Vhi$. Taking also into
account \Eref{ekss} we infer that the inflationary setting based
on \eqs{vhi}{VJe} is reproduced.


\renewcommand{\arraystretch}{1.2}
\begin{table}[t!]
\bec\begin{tabular}{|c|c|c|c|} \hline
{\sc Fields}&{\sc Eingestates} & \multicolumn{2}{|c|}{\sc Masses
Squared}\\\hline\hline
\multicolumn{4}{|c|}{Scalars}\\ \hline
$1$ real &$\what \th$ & $m^2_{\th}$&$6\Hhi^2$\\
$1$ complex&$S$ &$m^2_{S}$&
$(6/\nst+3n^2\fm^{p+2}/\nm^2pN\sg^{\nm}\fr)\Hhi^2$\\
\hline%
\multicolumn{4}{|c|}{Spinors}\\\hline
$2$ Weyl &$({{\psi}_{S}\pm
\what{\psi}_{\Phi})/\sqrt{2}}$&${m}^2_{\psi\pm}$&
$m^2_{S}-6\Hhi^2/\nst$ \\\hline
\end{tabular}\eec
\caption{\sl\small Mass-squared spectrum along the path in Eq.
(3.5) for $K=\tklmst$ and \m=\e\ or \t.}\label{tab1}
\end{table}
\renewcommand{\arraystretch}{1.}

The presence of $S$ and $\th$ in our SUGRA embedding obliges us to
check if the configuration in \Eref{inftr} is stable w.r.t the
excitations of those fields. In particular, we find the
expressions of the masses squared $m^2_{\chi^\al}$ (with
$\chi^\al=\th$ and $S$) arranged in \Tref{tab1}. Thanks to the
parameter $\nst$ with $0<\nst<6$ -- in practise we use $\nst=1$
--, $m^2_{S}$ can be retained positive. Nonetheless $
m^2_{\chi^\al}\gg\Hhi^2=\Vhio/3$ for $\sgf\leq\sg\leq\sgx$ and so
the one-field inflationary setting remains intact. In \Tref{tab1}
we display the masses $m^2_{\psi^\pm}$ of the corresponding
fermions with $\psi_{S}$ and
$\what\psi_{\Phi}=\sqrt{K_{\Phi\Phi^*}}\psi_{\Phi}$ being the Weyl
spinors associated with $S$ and $\Phi$ respectively. Inserting the
derived mass spectrum in the well-known Coleman-Weinberg formula
we can find the one-loop radiative corrections, $\dV$ to $\Vhi$ --
cf. \cref{jhep, rcellis}. It can be verified \cite{actpole} that
our results are immune from $\dV$, provided that the
renormalization group mass scale $Q$, is determined by requiring
$\dV(\sgx)=0$ or $\dV(\sgf)=0$. Under these circumstances, our
results within SUGRA can be reproduced by using exclusively the
ingredients of \eqs{vhi}{VJe} of the non-SUSY set-up.

\section{Inflation Analysis} \label{ana}

The establishment of a period of slow-roll \etpmi\ is controlled
by the condition
\beq{\ftn\sf
max}\{\eph(\phi),|\ith(\phi)|\}\leq1,\label{srcon}\eeq
where the slow-roll parameter $\eph$ takes a common form for both
$\klm$ in \Eref{ket} which is
\beqs \beq \label{eph} \epsilon=\left({\Ve_{\rm
I,\se}\over\sqrt{2}\Ve_{\rm I}}\right)^2=\frac{n^2
\fm^{p+2}}{\nm^2p N\sg^{\nm}\fr}\>\>\mbox{for}\>\>K=\klm.\eeq
On the other hand, $\ith$ can be expressed by separate formulas
for $\kbe$ and $\kat$ as follows
\beq\label{eta} \eta={\Ve_{\rm I,\se\se}\over\Ve_{\rm I}}\simeq
\bcs {n\fe^{p+1}}(2n\fe-2-p\sg)/{pN\sg\far} &\mbox{for $K=\kbe$,} \\
n\fb^{p+1}\lf n
\fb\fbr-1-\sg^2-p\sg^2(p\sg^2+3)\rg/{2pN\fbr^2\sg^2} &\mbox{for
$K=\kat$.}\ecs \eeq\eeqs
We remark that the expressions of both $\eph$ and $\eta$ includes
$\fm$ in \Eref{fp} in the numerator and so, these can be kept low
enough for $\sg$ close to unity. Expanding $\fm$ for $\sg\ll1$ and
neglecting terms of order $\sg^{2\nm}$ or larger we obtain
$\sg_{\rm f}$ which saturates \Eref{srcon} and can be approximated
as \cite{actpole}
\beq \sg_{\rm f}\simeq\bcs 2n\lf n(2+p)+\sqrt{4Np(1+4p)+n^2(p+2)^2}\rg^{-1}&\mbox{for $K=\kae$,}\\
n/\sqrt{n^2 (2 + p)-4 N p}&\mbox{for $K=\kat$.}\ecs
\label{sgf}\eeq

The number of e-foldings $\Ns$ that the scale $\ks=0.05/{\rm Mpc}$
experiences during \etpmi\ can be computed using the standard
formula
\beq \Ns=\int_{\sef}^{\sex} d\se\frac{\Vhi}{\Ve_{\rm
I,\se}}=\bcs (N/2n)\lf1+({\fars-1})/{\fes^{p+1}}\rg &\mbox{for $K=\kbe$,}\\
Np\sgx^2/n{\fbs^{p+1}} &\mbox{for $K=\kat$,}\ecs \label{Nhi} \eeq
where $\sgx~[\sex]$ is the value of $\sg~[\se]$ when $\ks$ crosses
the inflationary horizon and we take in account that $\sgx\gg\sgf$
in the analytic expressions above. Hereafter, the variables with
subscript $\star$ are evaluated at $\sg=\sgx$. Note that both
expressions od $\Ns$ includes $\fm$ in \Eref{fp} in the
denominator and so for $\sgx$ values sufficiently close to unity
we are able to obtain a large enough value of $\Ns$. Setting,
therefore, $\sgx^{\nm}\simeq1$ in the numerators of the
expressions above we can simplify them and solve w.r.t $\sgx$ as
follows
\begin{equation}
\label{sgx} \Ns\simeq\bcs N(1+{p}{\fes^{p+1}})/2n &\mbox{for $K=\kbe$}\\
N{p}/n{\fbs^{p+1}} &\mbox{for $K=\kat$}\ecs
\>\>\Rightarrow\>\>\sgx\simeq \lf1 - \lf \frac{\nm pN}{2n
\Ns}\rg^{\frac{1}{p+1}}\rg^{2^{1-\nm}}
\end{equation}
for any $\klm$ in \Eref{ket}. Note that for $K=\kbe$ we further
assumed that $\Ns\gg N/n$, which is confirmed a posteriori. It is
clear that we need to set $\sgx<1$ by construction to obtain a
large enough $\Ns$. This fact is welcome from the point of view of
the effective theory since it assists us to stabilize our
inflationary scenario against higher order terms in $W$ and/or
\klm -- see \eqs{Wn}{ktot}. Taking also into account that
$\Vhi(\sgx)^{1/4}\leq1$ -- see \sFref{fig1}{b} --, we expect that
corrections from quantum gravity may be, in principle, under
control.

The amplitude $\As$ of the power spectrum of the curvature
perturbations generated by $\sg$ can be calculated at $\sg=\sgx$
as a function of $\ld$. With given $\As$ we can also derive $\ld$
as follows
\beq \label{lan} \sqrt{\As}= \frac{1}{2\sqrt{3}\pi} \;
\frac{\Ve_{\rm I}^{3/2}(\sex)}{|\Ve_{\rm
I,\se}(\sex)|}\>\>\Rightarrow\>\>
\ld\simeq2^{\frac{5-2\nm}{2}+\frac{n}{4}}\frac{\sqrt{3\As}n\pi\fms^{p/2+1}}{
\sqrt{pN\sgx^{n+\nm}\frs}}\>\>\mbox{for}\>\>K=\klm, \eeq
where $\sgx$ is given by \Eref{sgx}. We did not replace it in the
expression above to avoid the exposition of the lengthly final
result.

The remaining inflationary observables (i.e. \ns, its running
$\as$ and $r$) are calculated via the relations
\beqs\baq \label{ns} && \ns=\: 1-6\epsilon_\star\ +\
2\eta_\star\simeq 1-\frac{p+2}{(p+1)\Ns},\\
&&\label{as} \as =\:{2\over3}\left(4\eta_\star^2-(n_{\rm
s}-1)^2\right)-2\xi_\star\simeq-\frac{p+2}{(p+1)N^2_\star},\\&&
\label{rs} r=16\epsilon_\star\simeq
2^{{(p(4-\nm)+2)}/(p+1)}\frac{(pn^pN)^{1/(p+1)}}{(p+1)N_\star^{(p+2)/(p+1)}},
\eaq\eeqs
where $\xi={\Ve_{\rm I,\widehat\sg} \Ve_{\rm
I,\widehat\sg\widehat\sg\widehat\sg}/\Ve_{\rm I}^2}$ and the
approximate expressions are obtained by expanding the exact result
for all possible $\klm$ in powers of $1/\Ns$ and keeping the
lowest order term. The common expressions for $\eph\gg\eta$ and
$\sgx$ in \eqs{eph}{sgx} justify the ``unified'' results obtained
for $\ns$, $\as$  and $r$ for all possible $\klm$. Note that $\ns$
and $\as$ are independent of $N$, $n$ and $\nm$ at the lowest
order as in the case of \etmi\ \cite{alinde,polec}. Also the
dependence of $r$ on $N$ is similar to that obtained for \etmi\
whereas the one on $n$ and $\nm$ is very faint. These prominent
features of our outputs signal the attractor behavior of our
solutions which are influenced only by the parameters $N$ and $p$.

\section{Numerical Results}\label{num}

The analytic findings above can be verified numerically
confronting the quantities in \eqs{Nhi}{lan} with the
observational requirements \cite{actin}
\beq \Ns \simeq61.3+\frac{1-3w_{\rm rh}}{12(1+w_{\rm
rh})}\ln\frac{\pi^2g_{\rm rh*}\Trh^4}{30\Vhi(\sgf)}+
\frac14\ln{\Vhi(\sgx)^2\over g_{\rm
rh*}^{1/3}\Vhi(\sgf)}\>\>\>\mbox{and}\>\>\sqrt{\As}\simeq4.617\cdot10^{-5},\label{Prob}\eeq
where we assume that \etpmi\ is followed in turn by an oscillatory
phase with mean equation-of-state parameter $w_{\rm rh}$ --
$w_{\rm rh}\simeq0$ for $n=2$ and $w_{\rm rh}\simeq1/3$ for $n=4$
-- radiation and matter domination. Motivated by implementations
\cite{tmhi} of non-thermal leptogenesis, which may follow \etpmi,
we set $\Trh\simeq10^9~\GeV$ for the reheat temperature with
corresponding energy-density effective number of degrees of
freedom $g_{\rm rh*}=228.75$.

\begin{figure}[!t]\vspace*{-.18in}\centering\hspace*{-0.3cm}
\includegraphics[width=60mm,angle=-90]{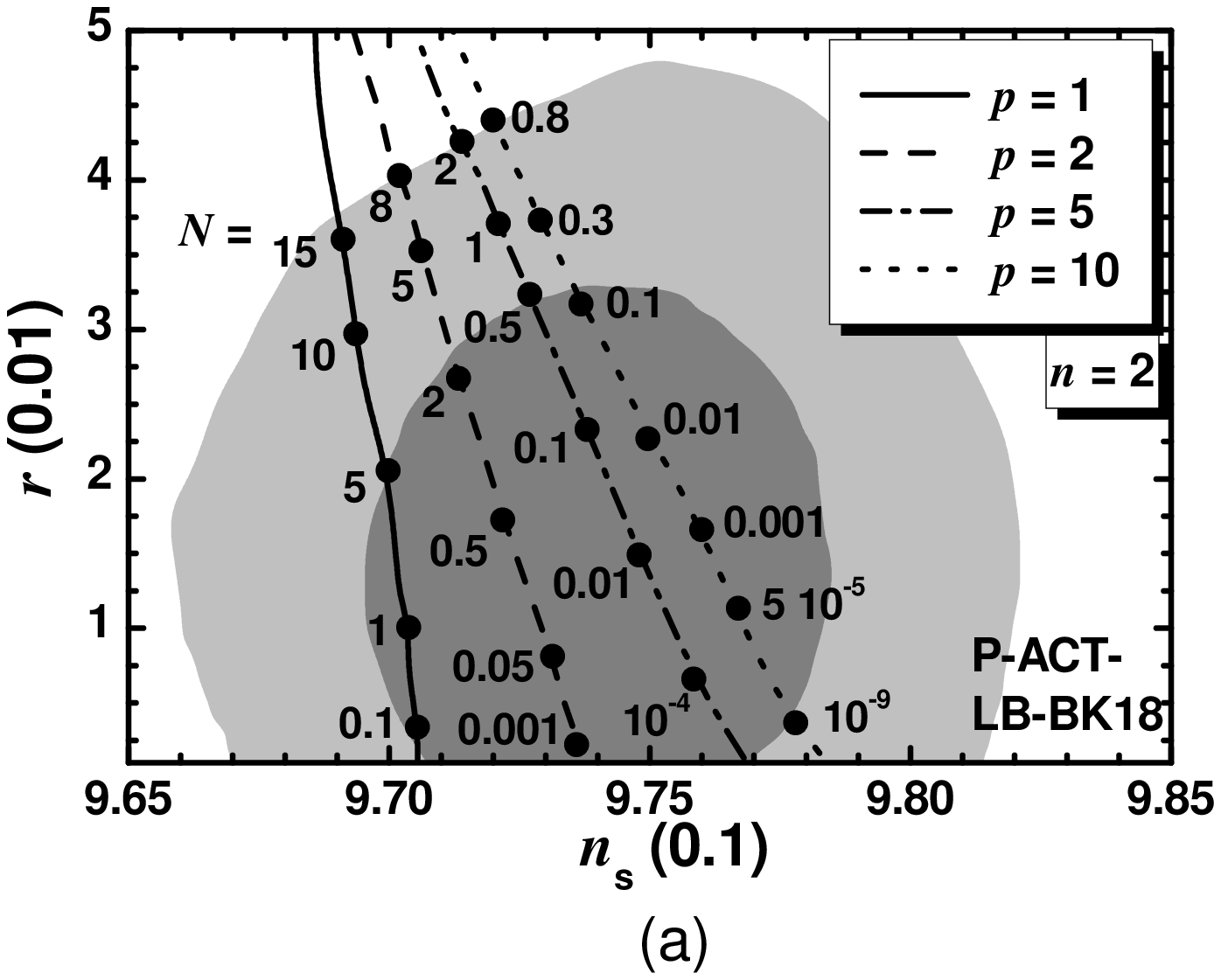}\hspace*{-0.7cm}
\includegraphics[width=60mm,angle=-90]{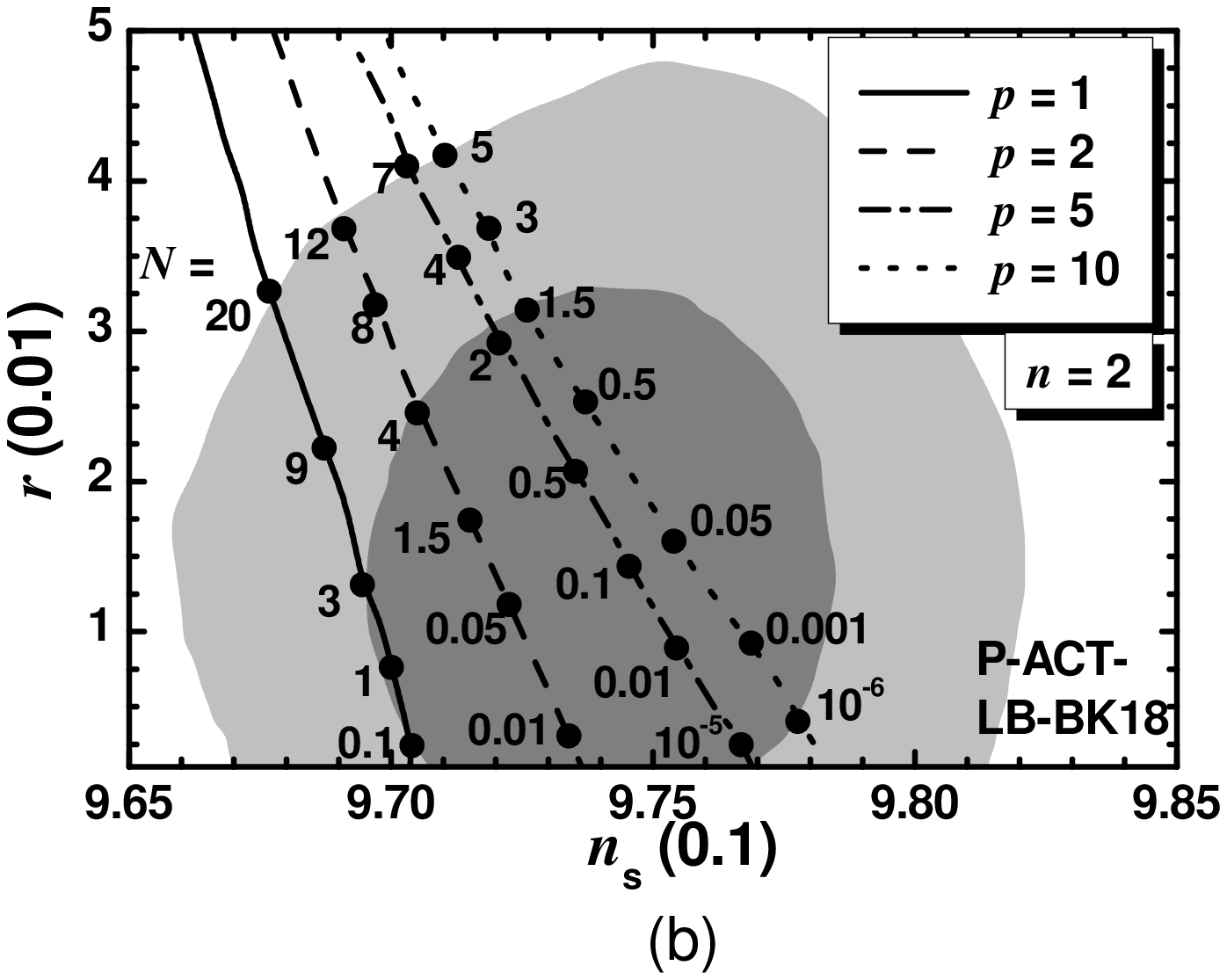}
\caption{\sl\small Allowed curves, as determined by Eq. (5.1), in
the $\ns-r$ plane for \epmi\ ({\sf\small a}) and \tpmi\
({\sf\small b}) with $n=2$, various $p$ values -- shown in the
legends -- and $N$ values indicated on the curves. The
marginalized joint $68\%$ [$95\%$] c.l. regions from \actc\ data
are also shown by the dark [light] shaded contours in the
background.}\label{fig2}
\end{figure}

\begin{figure}[!t]\vspace*{-.18in}\centering\hspace*{-0.3cm}
\includegraphics[width=60mm,angle=-90]{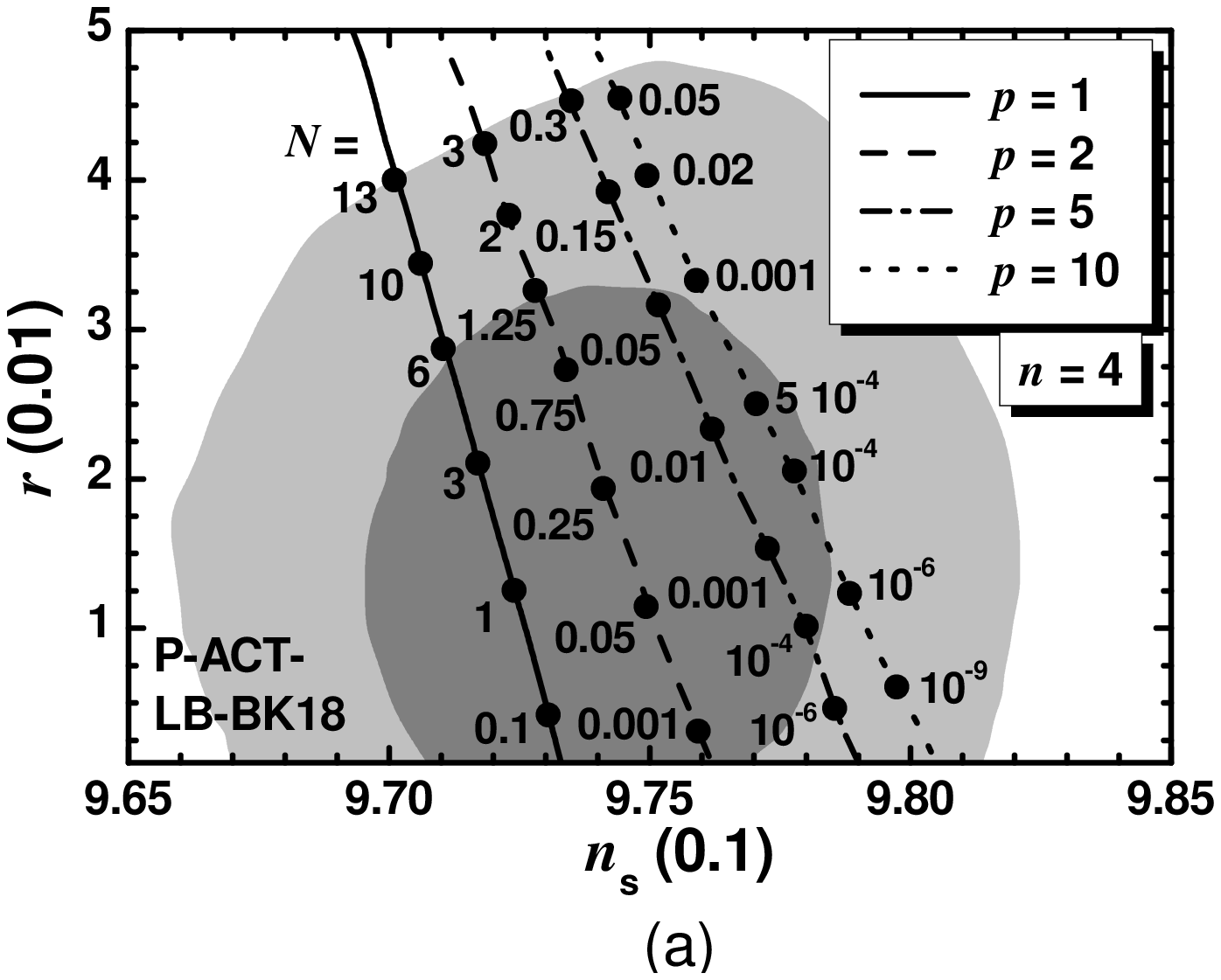}\hspace*{-0.7cm}
\includegraphics[width=60mm,angle=-90]{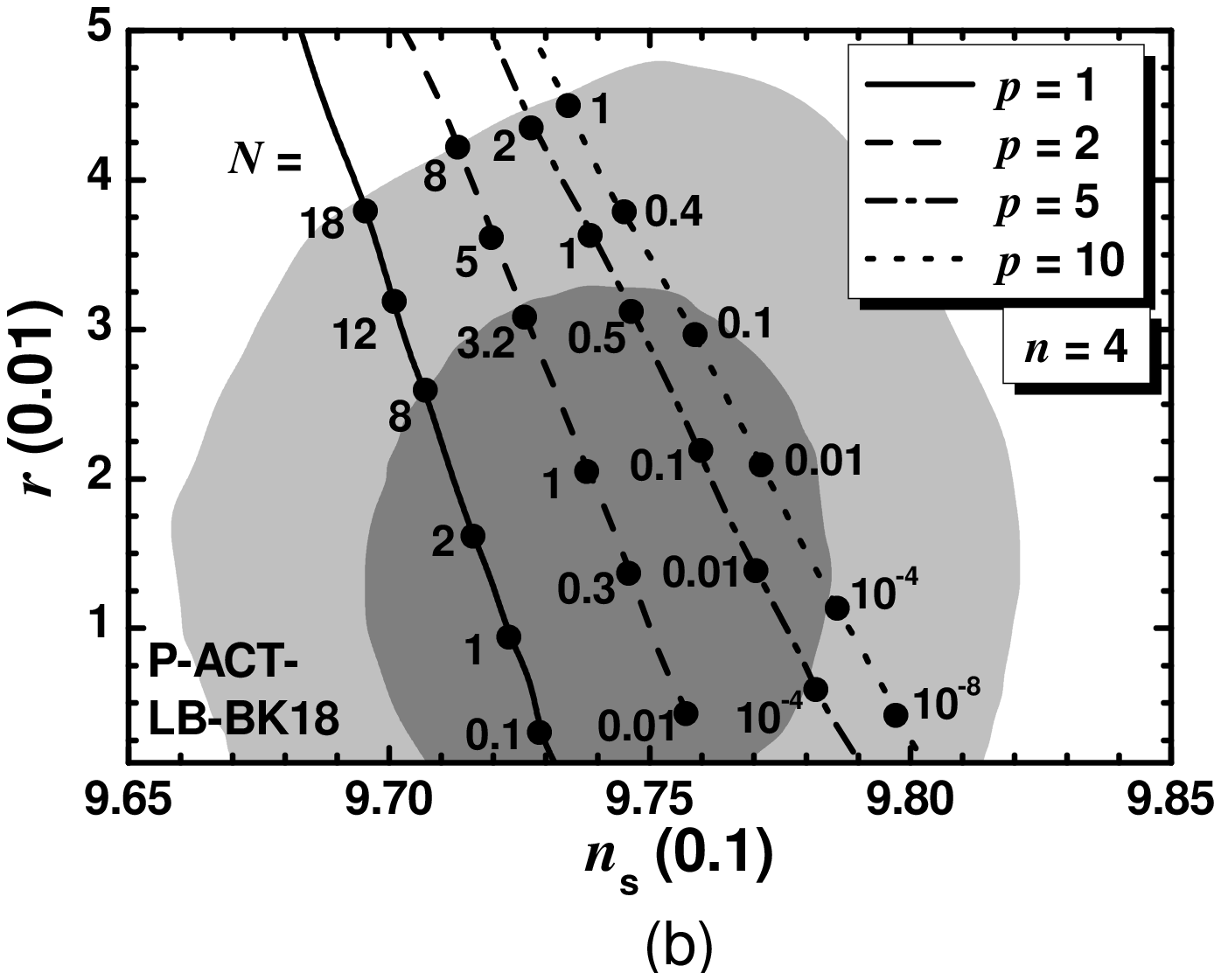}
\caption{\sl\small The same as in Fig.~2 but for
$n=4$.}\label{fig3}
\end{figure}

Enforcing \Eref{Prob} we can restrict $\ld$ and $\sgx$ as shown in
\eqs{Nhi}{lan}. Since the resulting $\sgx$ turns out to be close
to unity, a tuning of the initial conditions is required which can
be somehow quantified computing the quantity
\beq \Dex=\left(1 - \sgx\right).\label{dex}\eeq
The naturalness of the attainment of \etpmi\ increases with
$\Dex$. After that, we can estimate the models' predictions via
the definitive expressions in \eqss{ns}{as}{rs}, for any selected
set $(n,N,p)$ and compare them with the \actc\ data \cite{actin}
as done in \Fref{fig2} for $n=2$ and \Fref{fig3} for $n=4$. In
both cases our theoretical outputs are encoded as lines in the
$\ns-r$ plane superposed on the $68\%$ and $95\%$ c.l.
observationally allowed regions, depicted by the dark and light
shaded contour respectively. We draw solid, dashed, dot-dashed and
dotted lines for $p=1, 2, 5$ and $10$ respectively and show the
variation of $N$ along each line.

From the figures above it is evident that the whole
observationally favored range in the $\ns-r$ plane is covered
varying $p$ and $N$. In particular, as inferred from \Eref{ns},
$\ns$ increases with $p$ and renders \etpmi\ excellently
compatible with data. However, this increase becomes slower for
large $p$'s and so no upper bound on $p$ can be inferred by the
upper bound on $\ns$ in \Eref{data}. For this reason we set by
hand $p\leq10$. On the other hand, $r$ increases with $N$ and
assumes, for natural $N$ values, values which lie within the reach
of planned experiments \cite{det} aiming to discover primordial
gravitational waves.

Comparing the plots for fixed $n$, i.e., \sFref{fig2}{a} with
\sFref{fig2}{b} and \sFref{fig3}{a} with \sFref{fig3}{b}, we see
that the $N$ values for \epmi\ turn out to be suppressed compared
to the values used in \tpmi, especially for large $p$'s. As a
consequence, lower $N$ values are necessitated for the attainment
of observationally acceptable $r$ values according to \Eref{data}.
E.g., for $(n,p)=(2,2)$ we obtain
\beq 10^{-4}\lesssim N\lesssim8 \>\>\mbox{for \epmi\
and}\>\>4\cdot10^{-3}\lesssim N\lesssim12 \>\>\mbox{for
\tpmi.}\>\>\nonumber\eeq
This fact signals some tuning which can be avoided selecting low
$p$ values especially in \epmi.

Comparing the plots for the same \etpmi\ but different $n$, i.e.,
\sFref{fig2}{a} with \sFref{fig3}{a} and \sFref{fig2}{b} with
\sFref{fig3}{b}, it can be deduced that as $n$ increases the
allowed curves move to the right. This is due to the fact that
$\Ns$ in the leftmost condition in \Eref{Prob} increases with $n$
(from about $51$ to $57$) since $w_{\rm rh}$ increases too and so
the resulting $\ns$ increases in accordance with the approximate
result in \Eref{ns}. This effect is disadvantageous for $n=2$ but
beneficial for $n=4$. Therefore, we could say that \epmi\ fits
better with $n=2$ whereas \tpmi\ with $n=4$.

\begin{figure}[!t]\vspace*{-.18in} \centering\hspace*{-0.3cm}
\includegraphics[width=60mm,angle=-90]{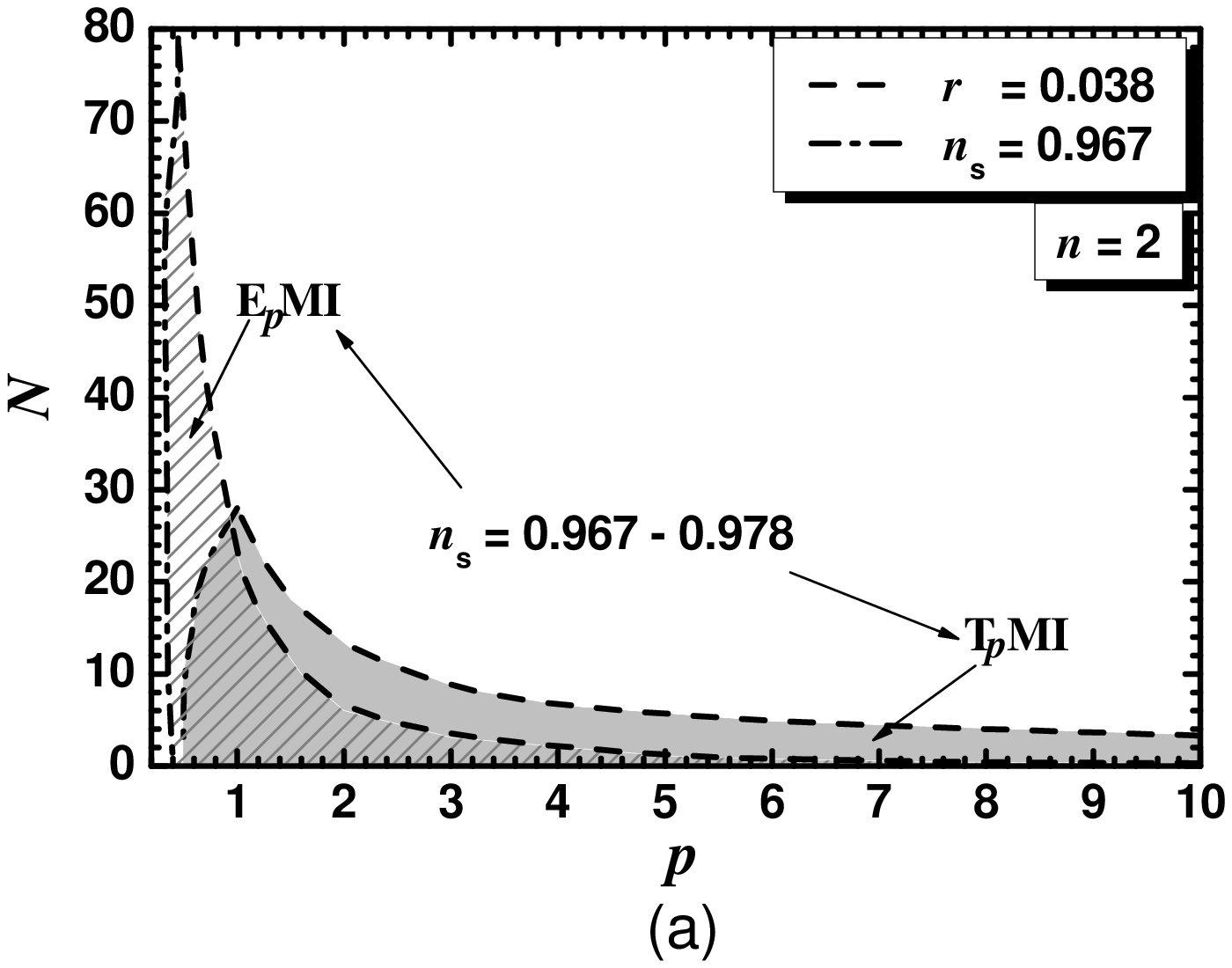}\hspace*{-0.7cm}
\includegraphics[width=60mm,angle=-90]{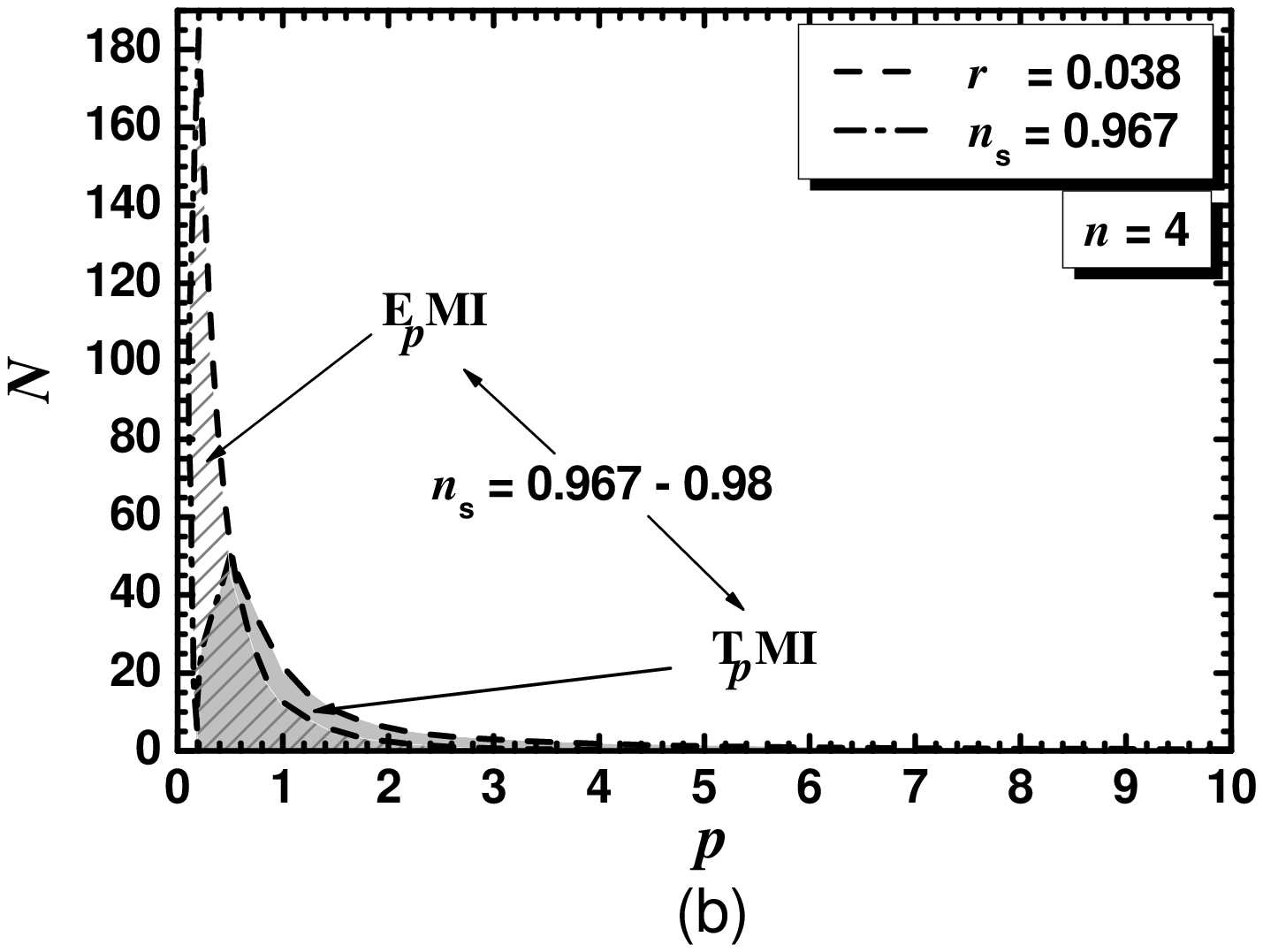}
\caption{\sl\small  Allowed (shaded for \tpmi\ and hatched for
\epmi) regions, as determined by Eq.~(1.3) and (5.1), in the $p-N$
plane for $n=2$ ({\sf\small a}) or $n=4$ ({\sf\small b}). The
conventions adopted for the boundary lines are also
shown.}\label{fig4}
\end{figure}

Our plots in \Fref{fig2} and \ref{fig3} together with \Eref{ns}
reveal that $\ns$ decreases with $p$ and so a lower bound on $p$
can be obtained from the lower bound on $\ns$ in \Eref{data}. On
the other hand, $N$ assumes a maximal value $N_{\rm max}$ from the
saturation of the bound on $r$ in \Eref{data}. Moreover, $N_{\rm
max}$ decreases as $p$ increases. The competition of both
restrictions on $N$ is shown in \sFref{fig4}{a} for $n=2$ and in
\sFref{fig4}{b} for $n=4$, where we delineate the allowed regions
(shaded for \tpmi\ and lined for \epmi) in $p-N$ plane. We see
that in both cases \epmi\ allows for larger $N_{\rm max}$ values
depending on $p$. Summarizing our results for $n=2$ -- see
\sFref{fig4}{a} -- we arrive at the following allowed ranges
\beqs\bea\label{res2} &&0.3\lesssim N_{\rm
max}\lesssim79\>\>\mbox{and}\>\>0.2\lesssim\Dex/100\lesssim61.4\>\>\mbox{for
\epmi,} \\&&3.3\lesssim N_{\rm
max}\lesssim28\>\>\mbox{and}\>\>1.6\lesssim\Dex/100\lesssim53.8\>\>\mbox{for
\tpmi.}\eea
Also $\as\simeq -(4.2-6.5)\cdot10^{-4}$ and $\Ns\simeq(50.5-52.1)$
for both \etpmi. On the other hand, for $n=4$ -- see
\sFref{fig4}{b} -- we obtain
\bea\label{res4} &&0.015\lesssim N_{\rm
max}\lesssim185\>\>\mbox{and}\>\>0.7\lesssim\Dex/100\lesssim45.3\>\>
\mbox{for \epmi,}\\ &&\>\>0.4\lesssim N_{\rm
max}\lesssim50\>\>\mbox{and}\>\>0.3\lesssim\Dex/100\lesssim39\>\>\mbox{for
\tpmi.}\eea\eeqs
Also $\as\simeq -(3.6-5.6)\cdot10^{-4}$ and $\Ns\simeq(55.6-57.9)$
for both \etpmi. It is notable that the maximal $\Dex$ values are
much larger than the ones derived within \etmi\ as those are
exposed in \cref{epole, sor} and therefore, the present models can
be characterized as more natural from the point of view of the
tuning in the initial conditions.

\section{Conclusions}\label{con}

We proposed a variant of \etmi\ (i.e., E- and T- model inflation)
named \etpmi\ using a gauge-singlet inflaton. In the non-SUSY
regime our setting was established as a non-linear sigma model
using chaotic potentials shown in \Eref{ci} and rational \Ka s
with a 1st or 2nd order pole and an overall exponent $p$ -- see
\Eref{ketp}. Within SUGRA we employ two chiral superfields, a
monomial superpotential, $W$, consistent with an $R$ symmetry --
see \Eref{Wn1} -- and a shift-symmetric rational $K$ for the
inflaton -- see \Eref{tks}. In both regimes our models develop an
attractor behavior -- i.e., an independence from $n$ and $\nm$,
e.g., in \eqs{Wn1}{kmsh} -- towards the \actc\ results for natural
$(p,N)$ values with possibly detectable primordial GWs. \tpmi\ can
be further realized by a gauge non-singlet chiral superfield as
shown in \cref{phi,phic}.

Despite the fact that our proposal employs one extra parameter
($p$) compared to the original model of \etmi\  and lacks any
string-theoretical motivation, the convergence of the results
towards the currently observational preferred values of \actc\
data is really impressive -- see \Fref{fig2} and \ref{fig3}.
Moreover, within our scheme, the reheating phase is not
constrained, as e.g. in \cref{att1, att3, laura, ellis10},
deformations of the simple chaotic potential is not applied, as
e.g. in \cref{ellis10, slinde}, and corrections to the
inflationary potential are not required, as e.g. in \cref{rcellis,
oxf, warm, waqas}.

\def\ijmp#1#2#3{{\emph{Int. Jour. Mod. Phys.}}
{\bf #1},~#3~(#2)}
\def\plb#1#2#3{{\emph{Phys. Lett.  B }}{\bf #1},~#3~(#2)}
\def\zpc#1#2#3{{Z. Phys. C }{\bf #1},~#3~(#2)}
\def\prl#1#2#3{{\emph{Phys. Rev. Lett.} }
{\bf #1},~#3~(#2)}
\def\rmp#1#2#3{{Rev. Mod. Phys.}
{\bf #1},~#3~(#2)}
\def\prep#1#2#3{\emph{Phys. Rep. }{\bf #1},~#3~(#2)}
\def\prd#1#2#3{{\emph{Phys. Rev.  D }}{\bf #1},~#3~(#2)}
\def\npb#1#2#3{{\emph{Nucl. Phys.} }{\bf B#1},~#3~(#2)}
\def\npps#1#2#3{{Nucl. Phys. B (Proc. Sup.)}
{\bf #1},~#3~(#2)}
\def\mpl#1#2#3{{Mod. Phys. Lett.}
{\bf #1},~#3~(#2)}
\def\arnps#1#2#3{{Annu. Rev. Nucl. Part. Sci.}
{\bf #1},~#3~(#2)}
\def\sjnp#1#2#3{{Sov. J. Nucl. Phys.}
{\bf #1},~#3~(#2)}
\def\jetp#1#2#3{{JETP Lett. }{\bf #1},~#3~(#2)}
\def\app#1#2#3{{Acta Phys. Polon.}
{\bf #1},~#3~(#2)}
\def\rnc#1#2#3{{Riv. Nuovo Cim.}
{\bf #1},~#3~(#2)}
\def\ap#1#2#3{{Ann. Phys. }{\bf #1},~#3~(#2)}
\def\ptp#1#2#3{{Prog. Theor. Phys.}
{\bf #1},~#3~(#2)}
\def\apjl#1#2#3{{Astrophys. J. Lett.}
{\bf #1},~#3~(#2)}
\def\n#1#2#3{{Nature }{\bf #1},~#3~(#2)}
\def\apj#1#2#3{{Astrophys. J.}
{\bf #1},~#3~(#2)}
\def\anj#1#2#3{{Astron. J. }{\bf #1},~#3~(#2)}
\def\mnras#1#2#3{{MNRAS }{\bf #1},~#3~(#2)}
\def\grg#1#2#3{{Gen. Rel. Grav.}
{\bf #1},~#3~(#2)}
\def\s#1#2#3{{Science }{\bf #1},~#3~(#2)}
\def\baas#1#2#3{{Bull. Am. Astron. Soc.}
{\bf #1},~#3~(#2)}
\def\ibid#1#2#3{{\it ibid. }{\bf #1},~#3~(#2)}
\def\cpc#1#2#3{{Comput. Phys. Commun.}
{\bf #1},~#3~(#2)}
\def\astp#1#2#3{{Astropart. Phys.}
{\bf #1},~#3~(#2)}
\def\epjc#1#2#3{\emph{Eur. Phys. J. C}
{\bf #1},~#3~(#2)}
\def\nima#1#2#3{{Nucl. Instrum. Meth. A}
{\bf #1},~#3~(#2)}
\def\jhep#1#2#3{{\emph{J. High
Energy Phys.} } {\bf #1},~#3~(#2)}
\def\jcap#1#2#3{\emph{J. Cosmol. Astropart.
Phys. } {\bf #1},~#3~(#2)}
\def\jcapn#1#2#3#4{\emph{J. Cosmol. Astropart.
Phys. }{\bf #1}, no. #4, #3 (#2)}
\def\prdn#1#2#3#4{{\sl Phys. Rev. D }{\bf #1}, no. #4, #3 (#2)}
\newcommand{\arxiv}[1]{{\small\tt  arXiv:#1}}
\newcommand{\hepph}[1]{{\small\tt  hep-ph/#1}}
\def\prdn#1#2#3#4{\emph{Phys. Rev. D }{\bf #1}, no. #4, #3 (#2)}
\def\jcapn#1#2#3#4{\emph{J. Cosmol. Astropart.
Phys. }{\bf #1}, no. #4, #3 (#2)}
\def\epjcn#1#2#3#4{\emph{Eur. Phys. J. C }{\bf #1}, no. #4, #3 (#2)}



\end{document}